\documentclass[aps,prc,preprint,superscriptaddress]{revtex4}

\usepackage[dvips]{graphicx}
\usepackage{dcolumn}
\usepackage{bm}
\usepackage{longtable}

\begin{document}

\preprint{Accepted for publication in Nature Physics (DOI:10.1038/NPHYS2457)}

\title{Production of spin-controlled rare isotope beams}

\author{Yuichi~Ichikawa}
\email[]{yuichikawa@phys.titech.ac.jp}
\altaffiliation{Present address: Department of Physics, Tokyo Institute of Technology,
  2-12-1 Oh-okayama, Meguro, Tokyo 152-8551, Japan}
\affiliation{RIKEN Nishina Center, 2-1 Hirosawa, Wako, Saitama 351-0198, Japan}
\author{Hideki~Ueno}
\affiliation{RIKEN Nishina Center, 2-1 Hirosawa, Wako, Saitama 351-0198, Japan}
\author{Yuji~Ishii}
\affiliation{Department of Physics, Tokyo Institute of Technology,
  2-12-1 Oh-okayama, Meguro, Tokyo 152-8551, Japan}
\author{Takeshi~Furukawa}
\affiliation{Department of Physics, Tokyo Metropolitan University,
  1-1 Minami-Ohsawa, Hachioji, Tokyo 192-0397, Japan}
\author{Akihiro~Yoshimi}
\affiliation{Research Core for Extreme Quantum World, Okayama University,
  3-1-1 Tsushimanaka, Kita, Okayama 700-8530, Japan}
\author{Daisuke~Kameda}
\affiliation{RIKEN Nishina Center, 2-1 Hirosawa, Wako, Saitama 351-0198, Japan}
\author{Hiroshi~Watanabe}
\affiliation{RIKEN Nishina Center, 2-1 Hirosawa, Wako, Saitama 351-0198, Japan}
\author{Nori~Aoi}
\affiliation{RIKEN Nishina Center, 2-1 Hirosawa, Wako, Saitama 351-0198, Japan}
\author{Koichiro~Asahi}
\affiliation{Department of Physics, Tokyo Institute of Technology,
  2-12-1 Oh-okayama, Meguro, Tokyo 152-8551, Japan}
\author{Dimiter~L.~Balabanski}
\affiliation{Institute for Nuclear Research and Nuclear Energy,
  Bulgarian Academy of Sciences, BG-1784, Sofia, Bulgaria}
\author{Rapha\"el~Chevrier}
\affiliation{CEA, DAM, DIF, F-91297 Arpajon, France}
\author{Jean-Michel~Daugas}
\affiliation{CEA, DAM, DIF, F-91297 Arpajon, France}
\author{Naoki~Fukuda}
\affiliation{RIKEN Nishina Center, 2-1 Hirosawa, Wako, Saitama 351-0198, Japan}
\author{Georgi~Georgiev}
\affiliation{CSNSM, IN2P3-CNRS, Universit\'e Paris-sud, F-91405 Orsay, France}
\author{Hironori~Hayashi}
\affiliation{Department of Physics, Tokyo Institute of Technology,
  2-12-1 Oh-okayama, Meguro, Tokyo 152-8551, Japan}
\author{Hiroaki~Iijima}
\affiliation{Department of Physics, Tokyo Institute of Technology,
  2-12-1 Oh-okayama, Meguro, Tokyo 152-8551, Japan}
\author{Naoto~Inabe}
\affiliation{RIKEN Nishina Center, 2-1 Hirosawa, Wako, Saitama 351-0198, Japan}
\author{Takeshi~Inoue}
\affiliation{Department of Physics, Tokyo Institute of Technology,
  2-12-1 Oh-okayama, Meguro, Tokyo 152-8551, Japan}
\author{Masayasu~Ishihara}
\affiliation{RIKEN Nishina Center, 2-1 Hirosawa, Wako, Saitama 351-0198, Japan}
\author{Toshiyuki~Kubo}
\affiliation{RIKEN Nishina Center, 2-1 Hirosawa, Wako, Saitama 351-0198, Japan}
\author{Tsubasa~Nanao}
\affiliation{Department of Physics, Tokyo Institute of Technology,
  2-12-1 Oh-okayama, Meguro, Tokyo 152-8551, Japan}
\author{Tetsuya~Ohnishi}
\affiliation{RIKEN Nishina Center, 2-1 Hirosawa, Wako, Saitama 351-0198, Japan}
\author{Kunifumi~Suzuki}
\affiliation{Department of Physics, Tokyo Institute of Technology,
  2-12-1 Oh-okayama, Meguro, Tokyo 152-8551, Japan}
\author{Masato~Tsuchiya}
\affiliation{Department of Physics, Tokyo Institute of Technology,
  2-12-1 Oh-okayama, Meguro, Tokyo 152-8551, Japan}
\author{Hiroyuki~Takeda}
\affiliation{RIKEN Nishina Center, 2-1 Hirosawa, Wako, Saitama 351-0198, Japan}
\author{Mustafa~M.~Rajabali}
\affiliation{Instituut voor Kern- en Stralingsfysica, K. U. Leuven,
  Celestijnenlaan 200D, B-3001 Leuven, Belgium}


\date{\today}

\begin{abstract}

The degree of freedom of spin in quantum systems serves as an unparalleled
laboratory where intriguing quantum physical properties can be observed,
and the ability to control spin is a powerful tool in physics research.
We propose a novel method for controlling spin in a system of rare isotopes
which takes advantage of the mechanism of the projectile fragmentation reaction combined with
the momentum-dispersion matching technique.
The present method was verified in an experiment at the RIKEN RI Beam Factory,
in which a degree of alignment of 8\% was achieved for the 
spin of a rare isotope $^{32}$Al.
The figure of merit for the present method was found to be greater than that
of the conventional method by a factor of more than 50.

\end{abstract}



\maketitle

The immense efforts expended to fully comprehend and control {\it quantum systems}
since their discovery are now entering an intriguing stage,
namely the controlling of the degree of freedom of spin~\cite{wolf,benioff,diamond,kato,hirayama}.
The case of nuclear systems is not an exception.
In recent years, nuclear physicists have been focusing their efforts
on expanding the domain of known species in the nuclear chart, which is
a two-dimensional map spanned by the axes of $N$ (number of neutrons) in the east direction
and $Z$ (number of protons) in the north direction.
The key technique used to explore the south eastern (neutron-rich, or negative in isospin $T_z$)
and north western (proton-rich, or positive in $T_z$) fronts of the map
has been the projectile fragmentation (PF) reaction,
in which an accelerated stable nucleus
is transmuted into an unstable one through abrasion upon collision with a target.
Several new facilities for providing rare-isotope (RI) beams by this technique,
such as RIBF~\cite{ribf} in Japan, FRIB~\cite{frib,frib-web1,frib-web2} in the
United States, and FAIR~\cite{fair1,fair2,fair-web} in Europe,
have been completed or designed for exploration of the frontiers of the nuclear chart.
Beyond such efforts toward exploring the $N$ and $Z$ axes, nuclear spin may be
a ``third axis'' to be pursued.
The study reported in the present article concerns the control of the
spin orientation of an unstable nucleus produced in a RI beam at such
fragmentation-based RI beam facilities.
The ability to control spin, when applied to state-of-the-art RI beams,
is expected to provide unprecedented opportunities for research on
nuclear structure of species situated outside the traditional region of the nuclear chart,
as well as for application in
materials sciences, where spin-controlled radioactive nuclei implanted in a sample could
serve as probes for investigating the structure and dynamics of
condensed matter~\cite{19F-review, diamond2}.

\begin{figure}[H]
  \begin{center}
	\includegraphics[width=12.cm]{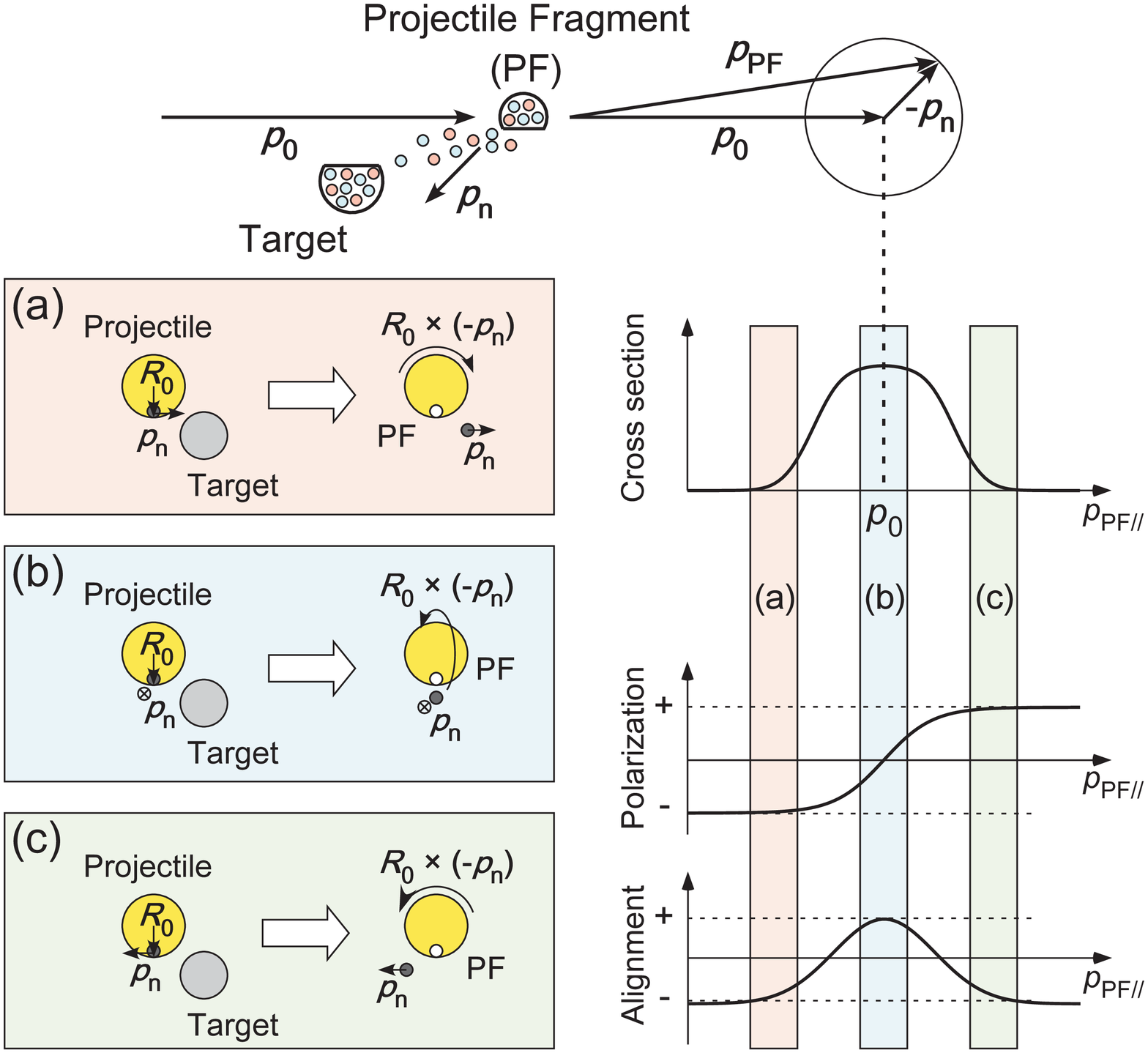}
  \end{center}
  \caption{Principle of producing spin orientation in PF reaction.
    A projectile with an initial momentum ${\boldsymbol p}_0$ is incident to the target.
    In a ``participant-spectator'' model,
    the nuclear spin ${\boldsymbol I}_{\rm {PF}}$ of the fragment arises
    from the angular momentum
    ${\boldsymbol R}_0 \times (-{\boldsymbol p}_{\rm n})$ 
    with respect to the center of mass of the projectile nucleus,
    owing to the removal of nucleons that, before collision, were in internal motion
    in the projectile nucleus (Fermi motion).
    Here, ${\boldsymbol p}_{\rm n}$ is the sum of momenta of the removed nucleons (participants),
    and ${\boldsymbol R}_0$ is the position vector of the participants in the projectile rest frame.
    Furthermore, the linear momentum ${\boldsymbol p}_{\rm {PF}}$ of the fragment is given as
    ${\boldsymbol p}_{\rm {PF}} = {\boldsymbol p}_0 - {\boldsymbol p}_{\rm n}$.
    Thus the orientation of the spin
    ${\boldsymbol I}_{\rm {PF}} = {\boldsymbol R}_0 \times
    ({\boldsymbol p}_0 -{\boldsymbol p}_{\rm {PF}})$
    is determined plainly as a function of the momentum
    ${\boldsymbol p}_{\rm {PF}}$ of outgoing fragment.
    In the figure, the axis of polarization is perpendicular to the reaction plane,
    while the axis of alignment is parallel to the beam axis.
    Insets (a), (b) and (c) illustrate cases which produce spin orientation for the fragments
    in the left wing, center and right wing of the momentum distribution, respectively. 
}
	  {\label{fig_pf}}
\end{figure}

The fragmentation of a projectile nucleus in high-energy nucleus-nucleus collisions is
described remarkably well by a simple model that assumes the projectile
fragment produced in the PF reaction to be a mere ``spectator'' of the projectile nucleus;
as a spectator, this fragment survives frequent nucleon-nucleon interactions,
and the other nucleons (``participants'') are abraded off through the reaction~\cite{hufner},
as illustrated in Fig.~\ref{fig_pf}.
In the model, the projectile fragment acquires an angular momentum (in other words, a nuclear spin),
whose orientation is determined simply as a function of the momentum of the outgoing fragment.
Although the spin orientation may practically be reduced due to cascade $\gamma$ decays following
the fragmentation, we assume that a significant amount of spin orientation survives in the fragment.
Here the degrees of spin orientation of rank one and two, in particular, are referred to as
{\it spin polarization} and {\it spin alignment}, respectively.
This implies a unique relation between the spin orientation and the direction of the removed momentum
${\boldsymbol p}_{\rm n}$, as illustrated in Fig.~\ref{fig_pf},
which can be utilized as an obvious means for producing spin-oriented RI beams.
One advantage of this method of orienting the fragment spin is that the resulting spin orientation
does not depend on the chemical or atomic properties of the RI.
However, the method also shows a drawback in the sense that the spin orientation thus produced in the PF
reaction tends to be partially or completely attenuated
since the fragmentation generally involves the removal of a large number of nucleons from the projectile.
This is quite a non-negligible flaw with respect to the yields attainable for spin-oriented beams
since high-intensity primary beams are only available for a limited set of nuclear species,
and consequently in most cases RIs of interest must be produced through the removal of a large number
of nucleons from the projectile.
Accordingly, there has been high demand for a new technique for preventing the attenuation in spin orientation
caused by large differences in mass between the projectile and the fragment.
In this paper, we present a method for producing highly spin-aligned RI beams by employing
a two-step PF process in combination with the momentum-dispersion matching technique.

\begin{figure}[p]
  \begin{center}
	\includegraphics[width=16cm]{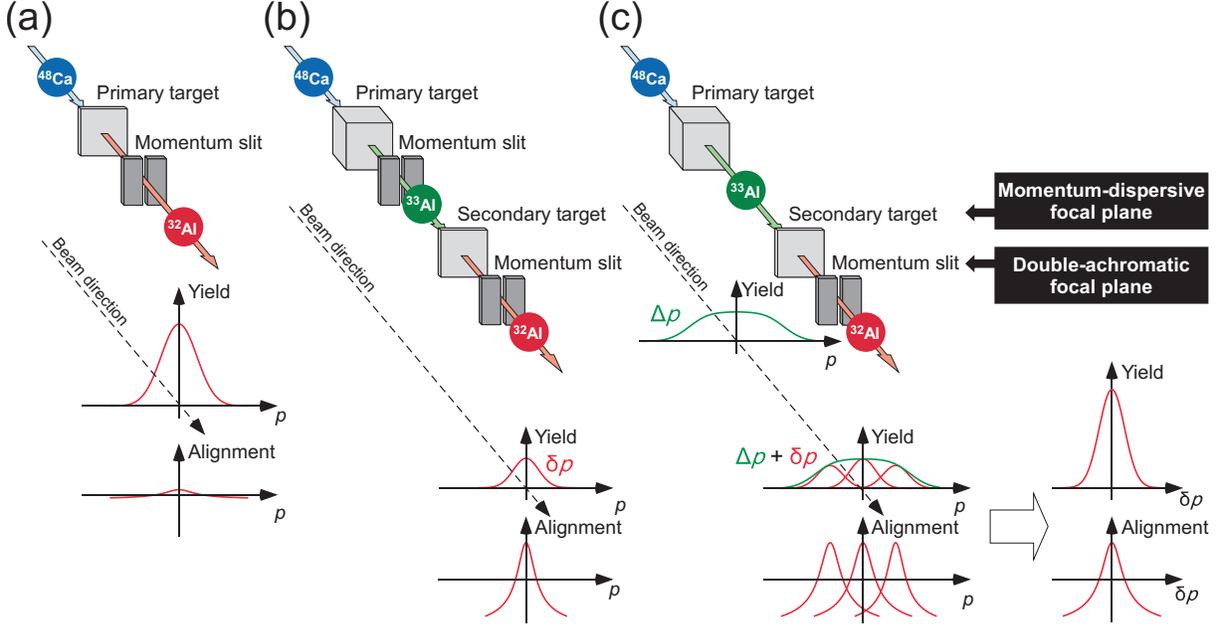}
  \end{center}
  \caption{Comparison of three schemes for producing a spin-aligned RI beam of $^{32}$Al
    from a primary beam of $^{48}$Ca.
    The graphs below each scheme represent the typical momentum distribution and the
    corresponding alignment, with abscissas representing the momentum $p$ of $^{32}$Al.
    (a) Single-step PF method.
    The $^{32}$Al beam is directly produced from $^{48}$Ca.
    Since PF involves a large number of nucleons, the expected spin alignment is small.
    (b) Two-step PF method.
    $^{32}$Al is produced via an intermediate nucleus $^{33}$Al.
    The expected spin alignment is high, whereas the production yield is low
    due to the two fold selection with momentum slits.
    (c) Two-step PF method with dispersion matching.
    Direct selection of the change in momentum $\delta p$ in the second PF can be achieved
    by placing a secondary target in the momentum-dispersive focal plane and a slit
    in the double-achromatic focal plane,
    because the momentum spread $\Delta p$ of the incident beam is compensated for
    by fulfilling the condition of momentum-dispersion matching.
    This method yields an intensive spin-aligned RI beam while avoiding cancellation
    between the opposite signs of spin alignment caused by the momentum spread $\Delta p$.
}
	  {\label{fig_2step}}
\end{figure}
 
\begin{figure}[p]
  \begin{center}
	\includegraphics[width=16cm]{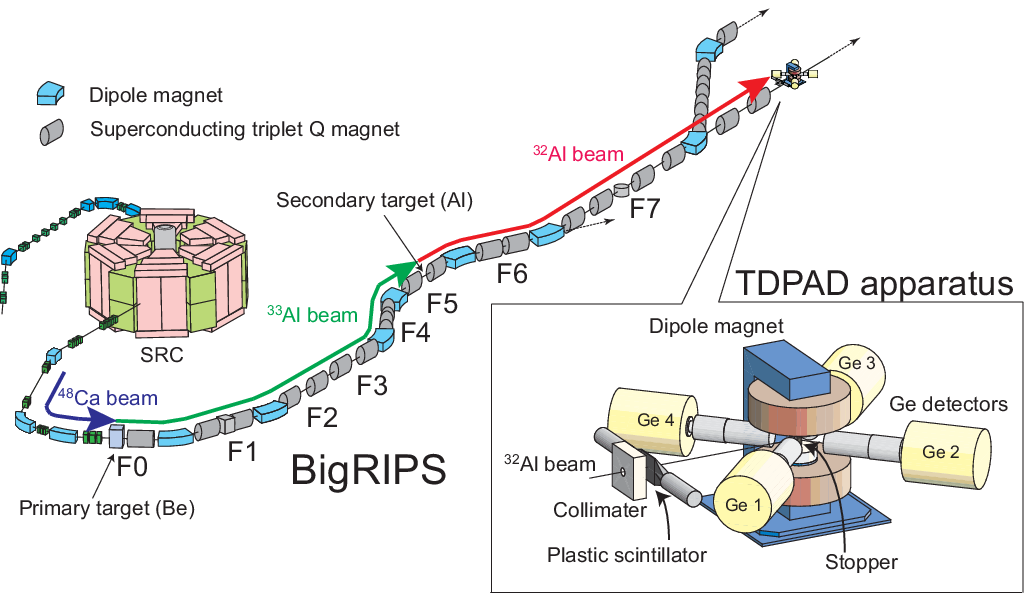}
  \end{center}
  \caption{Experimental setup.
    A primary beam of $^{48}$Ca ions,
    preaccelerated to 114~MeV/nucleon, is introduced into SRC (superconducting ring cyclotron),
    where it is further accelerated to 345~MeV/nucleon.
    The beam is subsequently adjusted to be incident to a primary target located in the focal plane F0.
    The projectile fragments are analyzed and selected in the BigRIPS beamline,
    where F1, F4, F5 and F6 are the momentum-dispersive focal planes
    and F2, F3 and F7 are the double-achromatic focal planes.
    In the present experiment, a second PF producing $^{32}$Al from $^{33}$Al takes place in F5.
    The inset shows the apparatus used for the TDPAD experiment.
  }
	  {\label{fig_exp}}
\end{figure}

Figure \ref{fig_2step} illustrates three different schemes for producing spin-aligned RI beams,
where each scheme uses a different configuration of elements,
namely primary and secondary targets and slits for selection.
The most basic scheme employs the configuration in (a), in which a nucleus of interest is
directly produced from a primary beam through a single occurrence of the PF (a single-step PF reaction).
As stated earlier, this scheme suffers from the drawback that the degree of spin alignment
tends to be attenuated when the PF involves the removal of a large number of nucleons from the projectile.
With the aim to overcome this problem, configuration (b) adopts a two-step PF reaction, where a beam of nuclei
produced in the first PF reaction (secondary beam) is used to obtain a beam of the nuclide of interest
through a second PF reaction.
In particular, using a slit installed at a momentum-dispersive focal plane,
the particles forming the secondary beam are chosen to be of a nuclide containing
one proton or neutron more than the nuclide of interest.
Thus, the target RI beam is produced via a PF reaction in which only one proton or neutron is removed.
For the RI beam obtained with this scheme, the spin alignment is expected to be high
due to the simplicity of the reaction~\cite{asahi-pol,ishihara-texas}.
We also note that a significant increase in the total production yield is suggested by
experiments~\cite{komeda} in in-beam $\gamma$-ray spectroscopy.
In the scheme presented above, however, the production yield is typically reduced
by a factor of $\sim$1/1000 because the production of target nuclei requires the successive occurrence of
two highly particular reactions.

A hint on how to eliminate the disadvantage of the scheme in (b) emerged from the recognition that
the quantity that determines the spin alignment is solely the momentum change $\delta p$ in the process
of fragmentation that produces the nuclei of interest, and that the spin alignment is not
sensitive to the momentum of the incident nucleus.
In scheme (b), $\delta p$ is selected with the aid of two momentum slits, the first of which is
used to select the momentum of the secondary beam and the second slit determines the outgoing momentum
of the secondary PF.
The tremendous and unnecessary drop in yield
is avoided by discarding the two-fold selection and introducing
a single direct selection of the target $\delta p$ itself.
This is realized by placing a secondary target in the momentum-dispersive focal plane
and a slit in the double-achromatic focal plane,
as illustrated in Fig.~\ref{fig_2step}~(c).
The concept of realizing maximum spectral resolution in momentum loss
by compensating for the beam momentum spread of the incident beam, as executed here, is known as
{\it dispersion matching} in ion optics~\cite{Co59,Bl71}.
The important point of this technique is that the reaction products that acquire equal amounts of
momentum change upon the second fragmentation are focused onto a single physical location.
The application of this technique to PF-induced spin alignment can prevent
the cancellation of opposite signs of spin alignment caused by momentum spread,
which secondary beams unavoidably undergo.

The validity of scheme (c) was first tested with the in-flight superconducting
RI separator BigRIPS~\cite{bigrips} at the RIKEN RIBF facility~\cite{ribf}.
The arrangement for the production of spin-aligned RI beams with the present method
is shown in Fig.~\ref{fig_exp}.
In the reaction at the primary target position F0, $^{33}$Al was produced by a PF reaction of
a 345-MeV/nucleon $^{48}$Ca beam on a $^9$Be target with a thickness of 1.85 g/cm$^2$,
chosen to provide a maximum production yield for the secondary $^{33}$Al beam.
A wedge-shaped aluminium degrader with a mean thickness of 4.05~g/cm$^2$ was placed
at the first momentum-dispersive focal plane F1,
where the momentum acceptance at F1 was $\pm$3\%.
The secondary $^{33}$Al beam was introduced to a second wedge-shaped aluminium target
with a mean thickness of 2.70~g/cm$^2$, placed at the second momentum-dispersive focal plane F5.
The $^{32}$Al nuclei (including those in isomeric state $^{32m}$Al) were produced
through a PF reaction involving the removal of one neutron from $^{33}$Al.
The thickness of the secondary target was chosen such that the energy loss from the target
was comparable with the theoretical estimate for the width of the momentum distribution~\cite{goldhaber}
for single-nucleon removal.
In the present case, $\sigma_{\rm Goldhaber} = 90$~MeV/$c$, and
the momentum width for $^{32}$Al was measured to be $\sigma=80$~MeV/$c$ or 0.4\%. 
The $^{32}$Al beam was subsequently transported to focal plane F7
whereby the momentum dispersion between F5 to F7 was tuned to be with the same
magnitude and opposite sign as that from F0 to F5 (momentum matching),
effectively canceling out the momentum dispersion from the site of the first PF reaction to F7.
We note that an admixture of $^{32}$Al particles in the $^{33}$Al secondary beam was found to be
negligibly small, due to the difference in the magnetic rigidity.
Thus, the $^{32}$Al particles that were solely produced at F5 were transported to F7.
The slit at F7 was used to select a region of momentum change at the second PF as
$\delta p/p=\pm0.15$\% about the center of relative momentum distribution.
The $^{32}$Al beam was then introduced to an experimental apparatus, shown in the 
inset of Fig.~\ref{fig_exp}, for time-differential perturbed angular distribution (TDPAD)
measurements.
(See ``Methods'' for details.)

\begin{figure}[p]
  \begin{center}
	\includegraphics[width=12.cm]{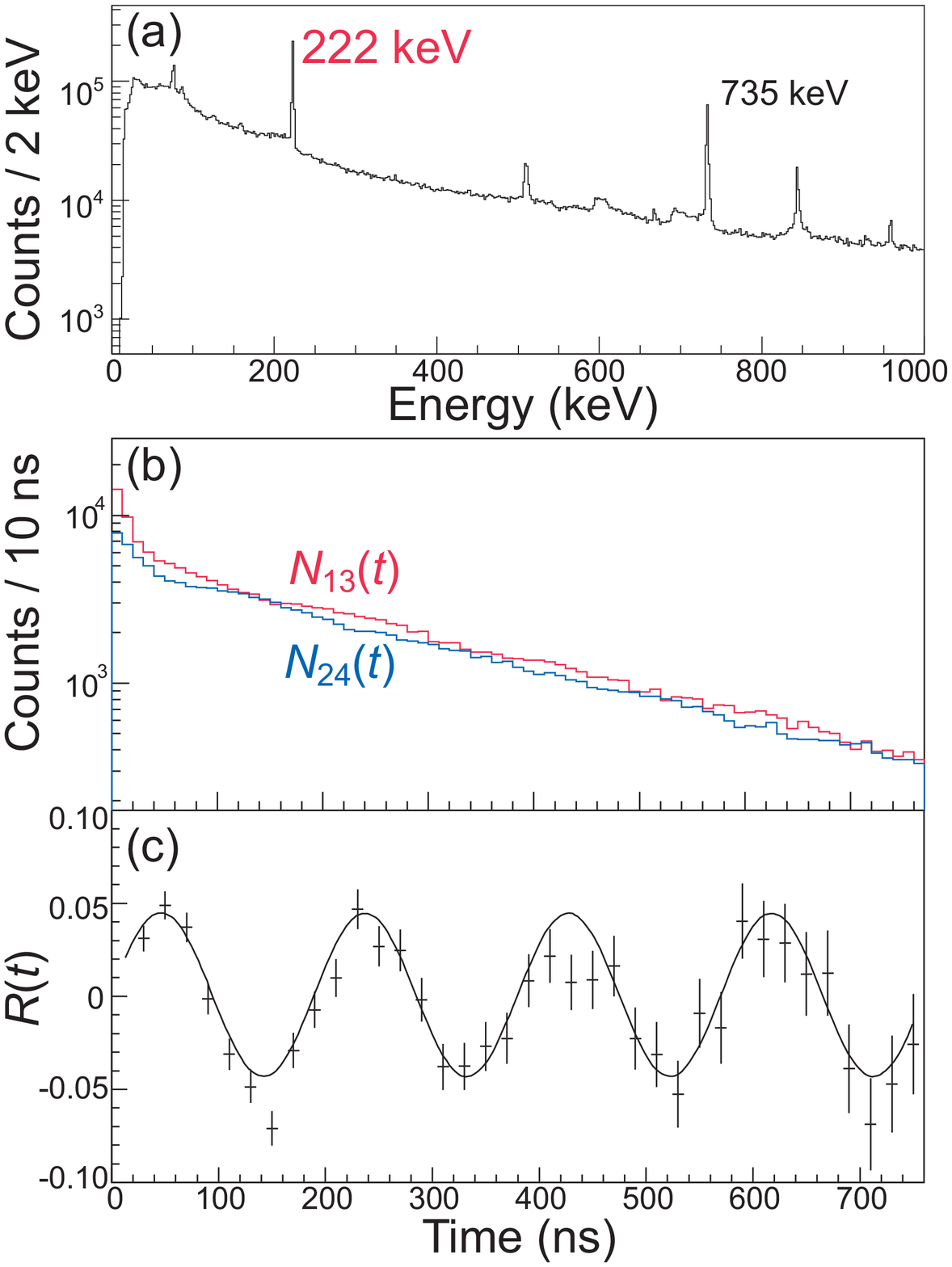}
  \end{center}
  \caption{Experimental results.
    (a)Energy spectrum. 
    De-excitation $\gamma$ rays emitted from $^{32m}$Al at 222~keV and 735~keV were observed.
    (b) Time variations of $N_{13}(t)$ and $N_{24}$(t) for 222-keV $\gamma$ rays.
    Events around $t=0$ originate from prompt $\gamma$ rays.
    (c) $R(t)$ ratio deduced from $N_{13}(t)$ and $N_{24}$(t), according to Eq.~\ref{eq_rfunc1}.
    The solid line represents the theoretical $R(t)$ function in Eq.~\ref{eq_rfunc2} after
    fitting to the experimental $R(t)$.
}
	  {\label{fig_rfunc}}
\end{figure}

The degree of spin alignment $A$ was determined from a ratio $R(t)$ defined as
\begin{equation}
  {\label{eq_rfunc1}}
  R(t) = \frac{N_{13}(t) - \epsilon N_{24}(t)}{N_{13}(t) + \epsilon N_{24}(t)},
\end{equation}
where $N_{13}$ ($N_{24}$) is the sum of the photo-peak count rates
at Ge 1 and Ge 3 (Ge 2 and Ge 4), which are two pairs of Ge detectors placed
diagonally to each other, as depicted in the inset of Fig.~\ref{fig_exp},
and $\epsilon$ denotes a correction factor for the detection efficiency.
Theoretically, the $R(t)$ ratio is expressed as a function of $t$ as 
\begin{equation}
  {\label{eq_rfunc2}}
 R(t) = \frac{3A_{22}}{4+A_{22}} \cos2(\omega_{\rm L} t + \alpha ),
\end{equation}
in terms of the rank-two anisotropy parameter $A_{22}$, which is defined as
$A_{22}=AB_2F_2$.
Terms with higher ranks were evaluated to be negligible in the
present case of $^{32m}$Al.
Here, $A$ denotes the degree of spin alignment
\begin{equation}
  {\label{eq_align}}
  A= \sum_m \frac{3m^2-I(I+1)}{I(2I-1)}a(m),
\end{equation}
where $a(m)$ is the occupation probability for magnetic sublevel $m$,
and $I$ the nuclear spin.
$B_2$ is the statistical tensor for complete alignment,
and $F_2$ is the radiation parameter~\cite{morinaga}.
The parameter $\omega_{\rm L}$ (Larmor frequency) is given by 
$ \omega_{\rm L} = g \mu_{\rm N} B_0 / \hbar$,
where $g$ is the $g$-factor of $^{32}$Al in units of the nuclear magneton
$\mu_{\rm N}$, and $\alpha$ is the initial phase of $R(t)$.

The $^{32}$Al nucleus is known to exhibit an isomeric state $^{32m}$Al~\cite{32al-isomer} 
at 957~keV with a half-life of 200(20)~ns.
The spin and parity of $^{32m}$Al have not been fixed
among the $4^+$ and $2^+$ candidates.
It is known that $^{32m}$Al undergoes de-excitation by E2 transition~\cite{32al-stephan}
with emission of $\gamma$ rays with an energy of 222~keV and subsequently decays
in cascade to the ground state by emitting 735-keV $\gamma$ rays.
Figure \ref{fig_rfunc}~(a) shows a $\gamma$-ray energy spectrum measured with the Ge detectors,
where 222-keV de-excitation $\gamma$ rays are clearly
observed as a peak.
The time variations $N_{13}(t)$ and $N_{24}(t)$ of the intensities for this peak
obtained with detectors pairs Ge 1~-~3 and Ge 2~-~4, respectively,
are presented in Fig.~\ref{fig_rfunc}~(b),
in which the corresponding abscissas represent the time difference of the signals at either
of the Ge detector pairs relative to the beam particle signal at a plastic scintillator placed
in front of the stopper crystal.
The $R(t)$ ratio evaluated according to Eq.~\ref{eq_rfunc1} is shown
in Fig.~\ref{fig_rfunc}~(c).

From the least $\chi ^2$ fitting of the theoretical function of Eq.~\ref{eq_rfunc2}
to the experimental $R(t)$ ratio of Eq.~\ref{eq_rfunc1},
we obtained the degree of spin alignment as $A=8(1)$\%, and
the $g$-factor of $^{32m}$Al was determined for the first time to be $g=1.32(1)$.
Also, the spin and parity were assigned to be $I^{\pi}=4^+$ through comparison
of the $g$-factor with theoretical calculations.
Detailed analysis and extended discussion regarding the $^{32}$Al nuclear structure based on
the obtained $g$-factor and spin-parity will be presented elsewhere.

A remeasurement of the degree of spin alignment was also performed during the experiment,
in which the momentum acceptance in the F5 focal plane was narrowed to be $\pm$0.5\%,
while maintaining other conditions unchanged.
This measurement corresponded to the two-step PF reaction without dispersion
matching (case (b) in Fig.~\ref{fig_2step}).
The degree of spin alignment derived from this measurement, 9(2)\%, 
is consistent with the above value obtained with the proposed method, 8(1)\%,
thus confirming that the present method of producing spin-aligned RI beams is
valid and performs well.

\begin{table}[p]
  \caption{Comparison of the two-step and the single-step PF methods.
    $\sigma$ and $\Delta p /p$ denote the width of the momentum distribution and the relative width
    of selected region around the center of the fragment momentum distribution, respectively.
    $Y(^{32}{\rm Al})$, $Y(^{32m}{\rm Al})$ and $p(^{32}{\rm Al})$ are
    the yield of $^{32}$Al beam particle, the yield of its isomeric state $^{32m}$Al,
    and the purity of $^{32}$Al particles in the RI beam, respectively, at
    the final focal plane for each method.
    The isomer to ground-state ratio $r$ for the production of $^{32}$Al was deduced
    as $r= Y(^{32m}{\rm Al}) / Y(^{32}{\rm Al})d$,
    where $d$ is the correction factor for the in-flight decays of $^{32m}$Al from
    the production target to the final focal plane,
    which are 0.47 and 0.17 for the two-step and single-step methods, respectively.
    $A$ is the degree of spin alignment. 
    The values of FOM are defined as proportional to $ A^2 \times Y(^{32m}{\rm Al})$
    and normalized so that the FOM for the two-step method is unity.
  }
 \label{tab_fom}
 \begin{center}
   \begin{tabular}{l   c c }
     &  Two-step method & Single-step method \\
     \hline
     Reaction & $^{48}$Ca $\to$ $^{33}$Al $\to$ $^{32}$Al & $^{48}$Ca $\to$ $^{32}$Al \\
     Energy & 200~MeV/nucleon & 345~MeV/nucleon\\
     Target  & 10-mm thick Be   &   4-mm thick Be \\
     $\sigma$  &  0.4\%  &  2.0\%\\
     $\Delta p/p$  &  $\pm$0.15\%  &  $\pm$0.5\%\\
     $p(^{32}{\rm Al})$ & 39(3)\% & 85(3)\%\\
     $Y(^{32}{\rm Al})$ & 2.3(2)~kcps & 8.6(3)~kcps (1/100 Att.)\\
     $Y(^{32m}{\rm Al})$ & 0.54(5)~kcps & 0.87(6)~kcps (1/100 Att.)\\
     $r$  &  50(6)\%  &   59(5)\%\\
     $A$  & 8(1)\%  & $<$0.8\%~(2$\sigma$)  \\
     FOM  & 1 & $< 0.02$ \\ 
     Meas. duration  & 11.9~h  & 9.3~h  \\
     \hline
   \end{tabular}
 \end{center}
\end{table}

A supplementary experiment was carried out in order to compare the performance of
the present method with that of the single-step method. 
$^{32}$Al was directly produced in a PF reaction of a $^{48}$Ca beam on a 4-mm thick Be target.
The thickness of the production target was chosen such that the energy loss
in the target was comparable with the Goldhaber width~\cite{goldhaber} (expected to be 4\% in this case).
In order to compare with the case of the two-step method under the equivalent condition, 
this measurement was carried out by selecting a momentum region of $\pm$0.5\% around the center of
fragment momentum distribution.
For this momentum region a maximum prolate alignment is expected.
As a result, the spin alignment was measured to be less than 0.8\% (2$\sigma$ confidence level).
A comparison of the two methods is summarized in Table~\ref{tab_fom}.
The figure of merit (FOM) for the production of such spin-aligned RI beams should be defined
to be proportional to the yield and the square of the degree of alignment.
In the measurement with the single-step PF reaction, a primary beam whose intensity was 
deliberately attenuated by a factor of 1/100 was used in order to avoid saturation in the
counting rate at the data acquisition system.
Here, the FOM was compared on the basis of actual effectiveness without correction for the attenuation,
in which the resulting FOM for the new method was found to be improved by a factor of more than 50.
Note that the degree of spin alignment in the single-step PF reaction could not be determined
within a measurement time comparable with that of the two-step PF reaction.
The superiority in FOM of the new method over the single-step PF reaction method should
be even more pronounced for nuclei located farther from the primary beam.

Theoretically, the maximum of the spin alignment for the case of single-nucleon removal from $^{33}$Al with
a momentum acceptance of $\pm 0.15$\% is estimated to be 30\% in a way similar to
that described in \cite{asahi-align,schmidt-ott}.
The estimation is based on a model proposed by H\"ufner and Nemes~\cite{hufner},
where the cross-section for the abrasion of one nucleon leading to a fragment of substate $m$ with momentum $p$
is proportional to the probability of finding a particle of substate $-m$ with momentum $-p$
at the surface of the target nucleus.
The maximum evaluated in this way is in fact four times greater than that obtained experimentally,
which may result from de-excitation from higher states populated through the PF reaction,
such as the ($4^-$)~\cite{32al-4minus} and $1^+$~\cite{32al-stephan} states.
This suggests that the ability to select the reaction path in populating the state of interest
is key to achieving augmented spin alignment.
Thus, spin alignment via PF reactions depends strongly
on both the reaction mechanism and the nuclear structure.
Under these circumstances, the achieved degree of alignment,
1/4 of the theoretical maximum which was obtained despite the situation that the reaction path
to the isomeric state was not unique, is rather satisfactory.
If we choose a nucleus produced by a unique reaction path,
a degree of spin alignment closer to the theoretical maximum might be possible to achieve.

\begin{figure}[p]
  \begin{center}
	\includegraphics[width=16.0cm]{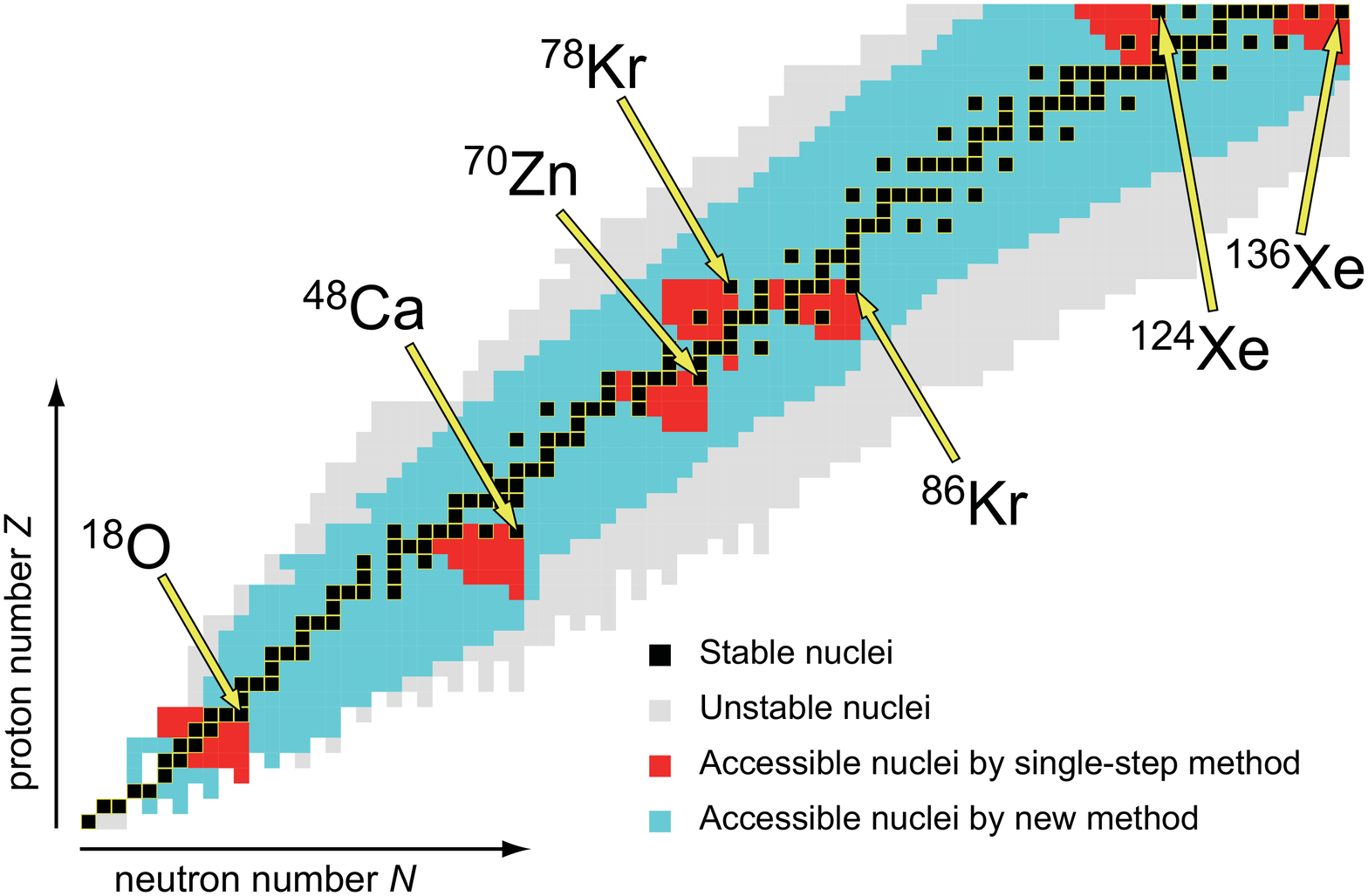}
  \end{center}
  \caption{Nuclear chart of ``accessible'' nuclei.
    Black boxes indicate stable nuclei, while colored boxes indicate unstable nuclei.
    Among the latter, red boxes represent those ``accessible'' with the single-step
    PF method, and blue boxes represent nuclei which are only accessible with the two-step PF method,
    where ``accessible'' here means that the nucleus of interest is producible with
    its spin aligned and with a production yield sufficiently large to determine
    the $g$-factor of its isomeric state with a 5~$\sigma$ confidence level in a one-day beam time.
    In the plot, primary beams are restricted to the typical beam particles which are available
    at high intensities at RIBF.
    Also, the following conditions are assumed:
    The degree of spin alignment is 10\% for single-nucleon removal from the beam particle,
    and reduces exponentially to 1\% down to 10-nucleon removal,
    as has been determined empirically;
    the values of intensity actually available at RIBF is assumed for each species of the primary beam;
    the cross-sections for the PF reactions are estimated based on parameter sets known as EPAX2~\cite{epax2},
    and the cross-section for the secondary PF reaction is assumed to be 1/1000, as usual; 
    the isomeric to ground state population ratio for the nucleus of interest in the PF reaction is 50\%;
    and the external magnetic field up to 1~Tesla is available for the TDPAD measurement.
  }
    	  {\label{fig_chart}}
\end{figure}

Figure \ref{fig_chart} shows the result of simulating the accessibility of unstable nuclei
via the two-step PF method (red region) and the conventional method (blue region).
In the simulation, the primary beam is assumed to be restricted to within a class of beam species
which are available with high intensities at RIBF~\cite{ribf}.
Clearly, the adoption of the two-step method drastically expands the set of accessible nuclei
in the nuclear chart.
In addition to a simplest case that the nucleus of interest was produced through the one-nucleon
removal reaction as presented in this article, the two-step scheme also allows to utilize few-nucleon
removal reactions as well as few-nucleon pickup reactions which are known to produce significant
spin orientation~\cite{pickup1}.

The FOM of our proposed method was found to be more than 50 times greater than that of the
conventional single-step PF reaction in this particular case and numerical simulations indicate
that the present method dramatically broadens the domain of accessible nuclei in the nuclear chart.
Such an ability to control spin, when applied to state-of-the-art RI beams, is expected to provide
unprecedented opportunities for research on the nuclear structure of species situated outside
the traditional region of the nuclear chart, as well as for applications in material research
where spin-controlled radioactive nuclei implanted in a sample serve as probes into
the structure and dynamics of condensed matter.

\section*{Methods}
Time-differential perturbed angular distribution (TDPAD) methods:
The $^{32}$Al beam was stopped in a Cu crystal stopper
mounted on the experimental apparatus for TDPAD measurements,
which was placed in a focal plane after the achromatic focal plane F7.
The TDPAD apparatus consists of a Cu crystal stopper, a dipole magnet, Ge detectors,
a plastic scintillator and a collimator, as shown in the inset of Fig.~\ref{fig_exp}.
The Cu stopper is 3.0 mm in thickness and 30$\times$30 mm$^2$ in area,
and the dipole magnet provides a static magnetic field $B_0=0.259$~T.
$^{32m}$Al are implanted into the Cu crystal, and de-excitation $\gamma$ rays are
detected with four Ge detectors located at a distance of 7.0~cm from the stopper
and at angles of $\pm45^{\circ}$ and $\pm135^{\circ}$ with respect to the beam axis.
The relative detection efficiency was 35\% for one and 15$-$20\% for the other three.
A plastic scintillator of 0.1 mm in thickness was placed upstream of the stopper,
the signal from which provided the time-zero trigger for the TDPAD measurement.
The TDPAD apparatus enabled us to determine the spin alignment as well as the $g$-factor
of $^{32m}$Al by observing the changes in anisotropy of the de-excitation
$\gamma$ rays emitted from spin-aligned $^{32m}$Al in synchronization
with the spin precession in the presence of an external magnetic field.

\begin{acknowledgments}
Experiments were performed under Program No.~NP0702-RIBF018 at RIBF,
operated by RIKEN Nishina Center and CNS, The University of Tokyo. 
We thank the RIKEN Ring Cyclotron staff for their cooperation during experiments.
Y.I. is grateful to the Special Postdoctoral Researchers Program, RIKEN.
This work was partly supported by the JSPS KAKENHI (22340071 and 20532089),
by the JSPS and MAEE under the Japan-France Integrated Action Program (SAKURA),
and by the Bulgarian National Science Fund (grant DID-02/16).
\end{acknowledgments}


\end{document}